\documentclass[twocolumn,showpacs,preprintnumbers,amsmath,amssymb]{revtex4}
\usepackage{epsfig}
\begin{document}

\title{Quantum Coherence between High Spin Superposition States of Single Molecule Magnet Ni$_4$}
\author{E. del Barco$^1$, A. D. Kent$^1$, E. C. Yang$^2$ and D. N. Hendrickson$^2$}
\affiliation{$^1$Department of Physics, New York University, 4
Washington Place, New York, NY 10003} \affiliation{$^2$Department
of Chemistry and Biochemistry, University of California San Diego
- La Jolla, CA 92093-0358}
\date{\today}

\begin{abstract}
Magnetic quantum tunneling in a single molecule magnet (SMM) has
been studied in experiments that combine microwave spectroscopy
with high sensitivity magnetic measurements. By monitoring
spin-state populations in the presence of microwave magnetic
fields, the energy splittings between low lying high spin
superposition states of SMM Ni$_4$ ($S$ = 4) have been measured.
Absorption linewidths give an upper bound on the rate of
decoherence. Pulsed microwave experiments provide a direct measure
of the spin-lattice relaxation time, which is found to be
remarkably long ($\sim$sec) and to increase with the energy
splitting.
\end{abstract}

 \maketitle

Magnetic quantum tunneling (MQT) in single molecule magnets
enables the creation of coherent high spin superposition states,
which are both of great fundamental interest and essential for the
use of SMMs in quantum computing \cite{Leggett,Leuenberger}. The
main focus of experiments conducted to date, however, have been on
incoherent MQT, where these superposition states are subject to
rapid decay into their classical counterparts: spin-up and
spin-down states \cite{Friedman,Wernsdorfer}. Decoherence
generally occurs when discrete levels are coupled to an
environment with many degrees of freedom, such as the modes of a
lattice (phonons), the electromagnetic field (photons) or nuclear
spins \cite{Prokofev,Stamp,Chudnovsky}. Here we investigate
coherent MQT in a SMM in which the tunneling rate is faster than
the rate of decoherence and at a temperature at which tunneling
occurs only between the lowest lying spin states. Our experiments
combine microwave excitations with high sensitivity magnetic
measurements and enable continuous monitoring of spin state
populations during the application of microwave radiation. Pulsed
microwave experiments provide a direct measure of the spin-lattice
relaxation time. Further, the linewidths associated with microwave
absorption provide an upper bound on the decoherence rate of a
high spin superposition state in a SMM.

We have chosen to study [Ni(hmp)(t-BuEtOH)Cl]$_4$, henceforth
referred to as Ni$_4$, because this is a particularly clean SMM,
with no nuclear spins on the transition metal sites \cite{Yang}.
The molecule core consists of four Ni$^{II}$ (Spin 1) magnetic
ions and oxygen atoms at alternating corners of a distorted cube,
with S$_4$ site symmetry. Ferromagnetic exchange interactions
between the Ni$^{II}$ ions lead to an $S=4$ ground state at low
temperature, as determined by magnetic susceptibility measurements
and high frequency Electron Paramagnetic Resonance (EPR)
\cite{Yang}. Magnetic hysteresis is observed below a
characteristic blocking temperature (1 K) and is associated with
the presence of a uniaxial magnetic anisotropy that favors high
spin projections on the easy axis of the molecule, which we denote
the z-axis. This uniaxial anisotropy leads to large energy barrier
($DS^2\sim12$ K) to magnetization reversal (see Fig. 1A).

MQT is characterized by the spin Hamiltonian:
\begin{equation}
\label{Eq.1}{\cal {H}}=-DS_z^2-\mu
_B\overrightarrow{S}\cdot\hat{g}\cdot\overrightarrow{H}\;
\end{equation}
where the first term is the uniaxial anisotropy and the second
term is the Zeeman energy associated with the interaction between
the spin and magnetic field. An external magnetic field applied
along the easy axis of the molecules $H_z$, tilts the double
potential well favoring those spin projections aligned with the
field. Note that in zero magnetic field the $|up\rangle$ and
$|down\rangle$ spin-projections have nearly the same energy and
MQT is possible. Importantly, in this case, a magnetic field
transverse to the anisotropy axis (in the $x-y$ plane) lifts the
degeneracy of these states by an energy $\Delta$, the tunnel
splitting, and leads to states that are coherent superpositions of
the original $|up\rangle$ and $|down\rangle$ spin-projections. The
frequency of MQT between $|up\rangle$ and $|down\rangle$ states is
proportional to the tunnel splitting ($f=\Delta/h$). Fig. 1C shows
the energy of the lowest lying states as a function of
longitudinal magnetic field. At zero field the states are
superpositions of opposite spin-projections
\begin{equation}
\label{Eq.2} \begin{array}{c}
  |S\rangle=\frac{1}{\sqrt{2}}(|up\rangle+|down\rangle) \\
  |A\rangle=\frac{1}{\sqrt{2}}(|up\rangle-|down\rangle)
\end{array}
\;
\end{equation}
where
\begin{equation}
\label{Eq.3} \begin{array}{c}
  |up\rangle=\frac{1}{\sqrt{2}}\sum_m(a_m+b_m)|m\rangle \\
  |down\rangle=-\frac{1}{\sqrt{2}}\sum_m(a_m-b_m)|m\rangle
\end{array}
\;
\end{equation}
The value of the coefficients $a_m$ and $b_m$ depend on the
applied field. Only the coefficients with $|m|=4$ are non-zero at
zero field and the two lowest levels are
$|S\rangle=\frac{1}{\sqrt{2}}(|+4\rangle+|-4\rangle)$ and
$|A\rangle=\frac{1}{\sqrt{2}}(|+4\rangle-|-4\rangle)$. In the
presence of a transverse magnetic field the $|up\rangle$ and
$|down\rangle$ states are tilted away from the $z$-axis but are
still separated by a large angle (see Figs. 1B and 1C), when this
field is less than the anisotropy field $H_a=2DS/(g\mu_B)=4.5$ T.
\begin{figure}
\begin{center}\includegraphics[width=8.5cm]{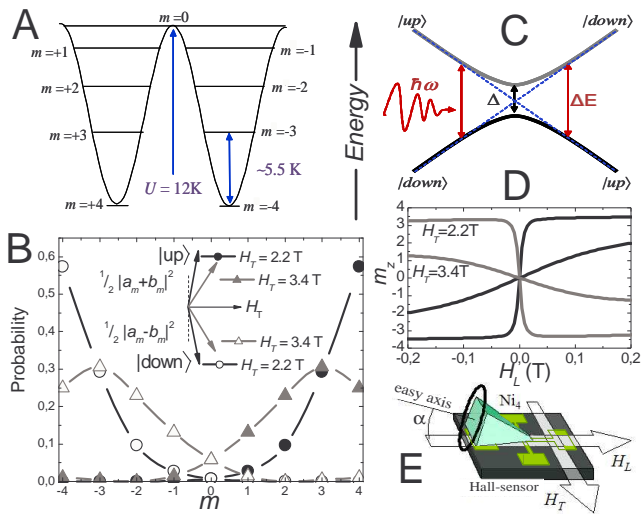}
\vspace{-3 mm} \caption{(Color on-line) A) Magnetic energy levels
of Ni$_4$, with different $z$-axis spin projections labeled. B)
The $|up\rangle$ and $|down\rangle$ state probability amplitudes
versus $z$-axis projection for different transverse magnetic
fields. C) Energy of the lowest lying levels in the vicinity of
zero field. D) Magnetization of lowest (black) and first excited
states (gray) as a function of the longitudinal field for two
different transverse fields. E) Experimental
configuration.}\vspace{-6 mm}
\end{center}
\end{figure}

The interaction of the spin with the environment limits the
coherence time of the superposition. A transverse field can be
used to increase the tunnel splitting and access a regime in which
the dynamics is expected to be coherent \cite{Stamp,Chudnovsky}.
It is straightforward to show that the tunnel splitting depends on
the transverse field to a high power, $\sim D(H_T/H_a)^{2S}$
\cite{Garanin}. This enables large variations in the magnitude of
the tunnel splitting for small changes in the transverse field.
Fig. 1D shows the $z$-component of magnetization for symmetric and
antisymmetric states as a function of the longitudinal magnetic
field. Photon induced transitions (PIT) between symmetric and
anti-symmetric states produce changes of the $z$ component of
magnetization when the longitudinal field is non-zero. When the
microwave field, $H_{ac}$, is parallel to the easy axis the
transition rate is given by the following expression:
\begin{equation}
\label{Eq.4}
\Gamma={\frac{\pi}{2}}\left({\frac{g\mu_B}{\hbar}}H_{ac}\right)^2
|\langle S|S_z|A\rangle|^2f(\omega)\;
\end{equation}
where $\langle S|S_z|A\rangle$ is the matrix element coupling the
symmetric and antisymmetric states. $f(\omega)$ is a Lorentzian
which characterizes the linewidth of the resonance:
\begin{equation}
\label{Eq.5}
f(\omega)={\frac{1}{\pi}}{\frac{\tau_2}{1+(\omega-\omega_0)^2\tau_2^2}}\;
\end{equation}
where $\omega_0=E/\hbar$, and $E$ is the energy level separation.
$\tau_2$ is the transverse relaxation time or decoherence time,
which sets the width of the resonance. Note that the transition
matrix element is maximum at zero magnetic field, when the states
are symmetric and antisymmetric superpositions of $|up\rangle$ and
$|down\rangle$ states. The matrix element decreases with
longitudinal magnetic field and approaches zero when
$2g\mu_BH>>\Delta$, and states are simple up and down spin
projections.

We have conducted experiments in a low temperature limit,
$k_BT<E$, in which there is a significant difference in the
population between the two lowest lying spin levels. A high
sensitivity micro-Hall magnetometer is used to measure the $z$
component of magnetization of a millimeter-sized single crystal of
Ni$_4$ that is placed with one of its faces parallel to the plane
of the Hall-sensor \cite{Kent} (see fig. 1E). The experiments were
conducted at 0.38 K in a He$^3$ cryostat that incorporates a 3D
vector superconducting magnet, in which magnetic fields can be
applied at arbitrary directions with respect to the axes of a
crystal. A thin circular superconducting loop ($\phi=2$ mm) is
placed with its plane perpendicular to the easy anisotropy axis of
the crystal (see Fig. 1E). This loop shorts the end of a 2.4 mm
coaxial line. A 50 GHz Vector Network Analyzer is used to apply
microwave radiation and enables characterization of the resonances
of the loop. All the measurements were performed at loop
resonances to maximize the power that is converted into ac
magnetic field at the sample position. Further, pulsed microwave
experiments were conducted using a pulse pattern generator to gate
the microwave source.

Figure 2A shows a magnetization curve recorded in the presence of
continuous-wave (CW) radiation at 39.8 GHz while a transverse
field of 3.2 T was applied. Peaks and dips are observed at
opposite polarities of the longitudinal field demonstrating PITs
between magnetic states of the molecules with opposite
spin-projections. Note that at PITs the sample temperature
increases by less than 0.01 K, showing that these features are not
due to sample heating. A careful inspection of the peaks shows a
more complex structure: each peak is formed by two peaks, labeled
$A$ and $B$. Measurements carried out at lower sweep rates show
that peak $B$ is also composed of two peaks, $B_1$ and $B_2$ (see
Figs. 3 and 4). This is in agreement with longitudinal field EPR
experiments, where different absorption lines have been found and
ascribed to molecules with slightly different values of the axial
anisotropy parameter \cite{Edwards}. The inset of Fig. 2A shows
the difference between the magnetization measured with and without
microwave radiation, $M-M_{eq}$, versus field, for frequencies
from 18 to 50 GHz and $H_T=3.2$ T.
\begin{figure}
\begin{center}\includegraphics[width=7.8cm]{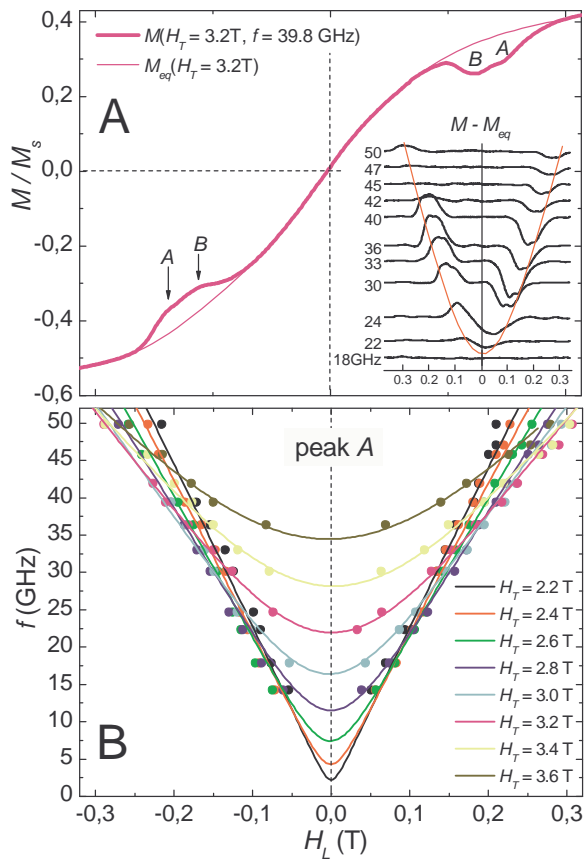}
\vspace{-3 mm} \caption{(Color on-line) A) Magnetization versus
longitudinal field in the presence of a transverse field,
$H_T=3.2$ T, while CW radiation of 39.8 GHz was applied. The inset
shows $M-M_{eq}$ versus longitudinal field at different microwave
frequencies. B) Field positions of the PITs of species $A$ for
several microwave frequencies and transverse fields.}\vspace{-8
mm}
\end{center}
\end{figure}

Classically the energy of these states would depend linearly on
magnetic field and simply cross at zero field. However, the
position of the peaks as a function of frequency does not depend
linearly on magnetic field, particularly near zero longitudinal
field. This is evident in Fig. 2B, which shows the position of
peak $A$ versus longitudinal field at different frequencies and as
a function of the transverse field (2.2 to 3.6 T). The curvature
of the energy splitting as the longitudinal field goes to zero is
evidence of level repulsion and, with the measured magnetization
changes, is a {\it clear signature of quantum superposition states
with opposite magnetizations}. The solid lines are fits of the
data by direct diagonalization of the Hamiltonian of Eq. (1) using
$D_A=0.765$ K, $B=7.9\times10^{-3}$ K, $C=3.25\times10^{-5}$ K,
$g_z=2.3$, $g_x=g_y=2.23$ \cite{note1}. The behavior of the
$B$-peaks can be fit with parameters $D_{B1}=0.735$ K and
$D_{B2}=0.745$ K (not shown). These values are in excellent
agreement with those determined from high frequency EPR
experiments \cite{Edwards}.
\begin{figure}
\begin{center}\includegraphics[width=7.5cm]{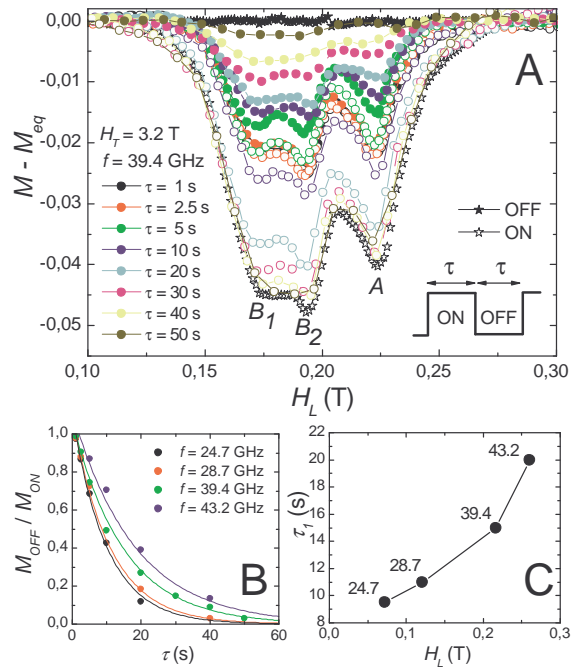}
\vspace{-3 mm} \caption{(Color on-line) A) Pulsed microwave
results: 50$\%$ duty cycle pulses of 39.4 GHz radiation are
applied while sweeping the longitudinal field across a PIT. The
transverse field is 3.2 T. The magnetization was recorded while
the pulse is ON (open circles) and OFF (close circles), for pulses
widths from 1 to 50 seconds. The black curves were measured with
99.99$\%$ (open black stars) and 0.01$\%$ (close black stars) duty
cycles. B) The ratio between $M_{OFF}$ and $M_{ON}$ for PIT $A$
versus pulse width for four different frequencies. The lines are
fits to an exponential decay with a characteristic time $\tau_1$.
C) $\tau_1$ as a function of the longitudinal field. A decrease of
the relaxation time is observed near zero field.}\vspace{-8 mm}
\end{center}
\end{figure}

In order to study the dynamics of the magnetization in the
presence of radiation, we have conducted experiments with pulsed
microwave radiation, in which the pulse time is of the order of
the longitudinal relaxation time ($\sim$sec). We sweep the applied
longitudinal magnetic field at a rate of $8.3\times10^{-5}$ T/s
with $H_T=3.2$ T around the position of a PIT. During this field
sweep we apply pulses of 39.4 GHz with a 50$\%$ duty cycle and
pulse times from 1 to 50 s. The results are shown in Fig. 3A. For
the smallest sweep rate used in this experiment, $8\times10^{-5}$
T/s, the field-width of the pulse is only 0.004 T (comparable to
the size of the dots). This indicates that inhomogeneous
broadening is not a significant component of the linewidth. The
ratio between the magnetization when the radiation is OFF (solid
circles) and ON (open circles) depends exponentially on the pulse
width (Fig. 3B), $M_{OFF}/M_{ON}=exp(-\tau/\tau_1)$. A fit to this
expression gives $\tau_1$, the longitudinal (or energy) relaxation
time, which governs the spontaneous decay of the excited state
population. We have repeated this experiment for frequencies
between 24.7 and 43.2 GHz at the same transverse field. The
corresponding relaxation curves are shown in Fig. 3B. $\tau_1$ is
plotted as a function of longitudinal field in Fig. 3C.
Remarkably, the relaxation time is long and increases from 8 to 20
s as the field and thus frequency increase. This is contrary to
general ideas that the relaxation time should decrease with
frequency, because of the increasing phase space available for
phonon generation \cite{note2,Prokofev2}.
\begin{figure}
\begin{center}\includegraphics[width=7.5cm]{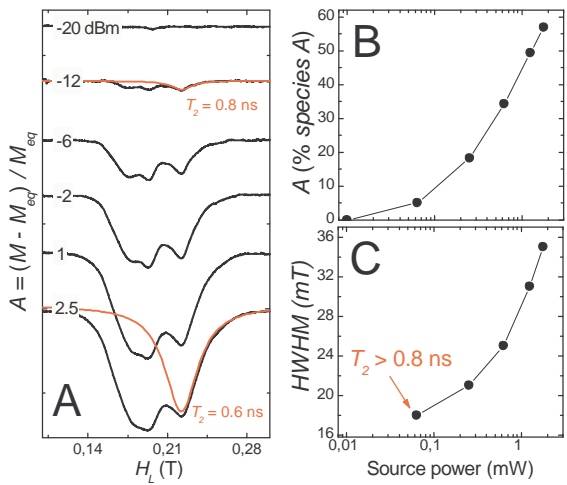}
\vspace{-3 mm} \caption{(Color on-line) A) $(M-M_{eq})/M_{eq}$ for
different powers with CW 39.4 GHz radiation applied to the sample.
The longitudinal field was swept at $1.6\times10^{-4}$ T/s and
$H_T=3.2$ T. Note that the curves are offset for clarity. The red
lines are fits to a Lorentzian with $\tau_2=0.6$ ns. Figures B and
C show the dependence of the amplitude and half width of PIT $A$,
respectively.}\vspace{-6 mm}
\end{center}
\end{figure}

We have measured the change in magnetic response as a function of
the power of the incident CW radiation (39.4 GHz) to estimate the
decoherence rate. A transverse magnetic field of 3.2 T is applied
and a longitudinal field is swept at a rate of $1.6\times10^{-4}$
T/s. The microwave power at the source was varied from -20 dBm to
2.5 dBm. The normalized change in magnetization
$(M-M_{eq})/M_{eq}$ is shown as a function of the longitudinal
field in Fig. 4A for different microwave powers. Fig 4B shows the
amplitude for peak $A$ versus power. The maximum power applied to
the sample produces transitions in 20$\%$ of the molecules
\cite{note3}. Larger magnetization changes (up to 30$\%$) have
been obtained with higher transverse fields and at lower microwave
frequencies (not shown). If we consider that this amplitude
corresponds to one of the peaks, which represents only a fraction
of the molecules in the crystal, then the amplitude for this
species is considerably larger. As an estimate, if each of the
species represented one third of the total, then a net amplitude
of 20$\%$ corresponds to 60$\%$ of molecules in that species. The
amplitude versus power for peak $A$ is shown in Fig. 4B (assuming
a 33$\%$ population). For this curve we can estimate the power
necessary to saturate the resonance. We find 50 mW (source power)
in this case (and about 5 mW with $H_T=3.4$ T and $f\sim30$ GHz,
not shown). We have fit the results shown in Fig. 4A to a
Lorentzian function $f(\omega)$ for the PIT $A$, as this feature
is easily distinguished from the others. We find that $\tau_2=0.6$
ns for the curve measured with the highest power (2.5 dBm) or a
resonance quality factor of $\sim10$.

In Fig. 4C we show the dependence of PIT $A$'s half width at half
maximum (HWHW) versus the microwave power at the source. The
minimum width (lowest power) allows us to estimate a lower bound
for the transverse relaxation time. The results from the fit of
the -12 dBm curve are shown in Fig. 4A and give $\tau_2>0.8$ ns.
The homogeneous linewidth represented by $\tau_2$ will be less
than that associated with inhomogeneous broadening. However, we
note that we recover the full amplitude of the peaks and dips
during the pulsed-radiation experiments (see Fig. 3A), for
sufficiently long pulses ($\tau>20$ s). This indicates that
$\tau_2\sim\tau_\phi$. A very similar decoherence time,
$\tau_\phi=1$ ns, has been recently reported in high frequency EPR
experiments carried out in a single crystal of a Mn$_4$ SMM dimer
($S=9/2$) \cite{Hill}.

This experiment demonstrates the vast difference in time scales
between the longitudinal and transverse relaxations times of high
spin superposition states. Clearly, for applications in quantum
computing decoherence is a major challenge, which is likely
amenable to synthetic strategies of nuclear isotope purification
and further reduction in intermolecular interactions.
Antiferromagnet molecular magnets with an uncompensated electronic
or nuclear spins that tracks the Neel vector for readout are also
excellent candidates for long coherence times in mesoscopic
spin-systems and for use in quantum computation \cite{Chiolero}.
These will be the subject of future investigations along with
studies of Rabi oscillations in molecular nanomagnets induced by
shorter and higher amplitude pulsed microwave magnetic fields.

This research was supported by NSF (Grant Nos. DMR-0103290,
0114142, and 0315609).


\begin{references}
\bibitem{note1} There is an additional transverse anisotropy term in eqn (1) of the
form C(S+4 + S-4) seen in EPR studies \cite{Yang,Edwards}.
\bibitem{note3}Considering the losses in the coaxial line ($\sim15$ dB)
and the measured reflected power ($\sim55$ dB), we estimate that
the power irradiated by the loop antenna is
$P_{loop}=0.032P_{source}$, (i.e. for $P_{source}=2.5$ dBm $=1.77$
mW the power radiated by the loop antenna is $P_{loop}\sim50 \mu$W
at 39.4 GHz). Taking into account the change in the magnetization
at the PIT, we estimate that the power absorbed by the sample is
2$\%$ of the power irradiated by the loop antenna. So for
$P_{loop}\sim50 \mu$W at 39.4 GHz, the power absorbed is $\sim1
\mu$W.
\end{references}
\end{document}